\documentclass[prl,twocolumn,showpacs,superscriptaddress,amsmath,amssymb]{revtex4}
\usepackage{graphicx}
\begin{document}
\title{Bistability and noise-enhanced  velocity of rolling motion}

\author{Thorsten P\"oschel}
\affiliation{Institut f\"ur Biochemie, Charit\'e, Monbijoustra{\ss}e 2, D-10117 Berlin, Germany}
\author{Nikolai V. Brilliantov}
\affiliation{Moscow State University, Physics Department, Moscow 119899, Russia}
\author{Alexei Zaikin}
\affiliation{Institut f\"ur Biochemie, Charit\'e, Monbijoustra{\ss}e 2, D-10117 Berlin, Germany}

\date{\today}
\begin{abstract}
  We investigate the motion of a hard cylinder rolling down a soft
  inclined plane. The cylinder is subjected to a viscous drag force
  and stochastic fluctuations due to the surrounding medium. In a wide
  range of parameters we observe bistability of the rolling velocity.
  In dependence on the parameters, increasing noise level may lead to
  increasing or decreasing average velocity of the cylinder.  The
  approximative analytical theory agrees with numerical results.
\end{abstract}
%46.30.Pa Friction, wear, adherence, hardness, mechanical contacts, tribology
%         (see also 81.40.P - in materials science)
%62.40.+i Anelasticity, internal friction, stress relaxation, and mechanical
%         resonances
%81.40.Pq Friction, lubrication, and wear (see also 46.30.P - in structural
%         mechanics)
\maketitle

Since 1785 when Vince described systematic experiments to determine
the nature of friction laws \cite{Vince:1785} the effect of rolling
friction has been investigated by many scientists according to its
great importance in engineering and natural sciences (e.g. \cite{many}
and many references therein).

In this letter we study the rolling motion of a hard cylinder on an
inclined soft plane.  Experimentally it was observed that the rolling
friction coefficient depends non-monotonously on the velocity
\cite{NonlinFrict}: For small velocities the rolling friction force
increases with increasing velocity, while for fast motion it decays as
the velocity grows. Furthermore, the cylinder is subjected to the drag
force of the surrounding medium, e.g. air, and unavoidably present
fluctuations. We will show that this system reveals non-trivial
dynamics, i.e., bistability of the cylinder's velocity and noise
controlled average rolling velocity.
%which implies an effective lubrication by noise.

{\em Model.} We consider a cylinder of radius $R$, mass $M$ and moment
of inertia $I$ which rolls at velocity $v$ on a plane \cite{foot}. The
cylinder is subjected to an external driving force $F_{\rm ex}=Mg\sin
\alpha$ due to the plane's inclination by the angle $\alpha$. The
rolling friction force $F_R$ and the viscous drag force $F_D$ due to
the surrounding air counteract this motion. We also assume that the
tangential force acting between the cylinder and the surface at the
contact area is strong enough to keep it from sliding. For certain
materials it has been shown that surface effects such as adhesion may
have significant influence on rolling friction (e.g. \cite{adhesion}).
On the other hand, for viscoelastic materials it was reported that
rolling friction is due very little to surface interactions, i.e., the
major part is due to deformation losses within the bulk of the
material \cite{noadhesion}. Here we assume that surface effects may be
neglected.

Hence, Newton's  equation for the cylinder reads
\begin{equation}
\label{Newton} \left(M +I/R\right) \dot{v} = -F_D(v)-F_R(v)+F_{\rm
ex} +\zeta(t) \, .
\end{equation}
The stochastic force  $\zeta(t)$ describing fluctuations in the
media is modeled  by Gaussian white noise of zero average,
$\langle \zeta(t)\rangle=0$, and intensity $\sigma$: $\langle
\zeta(t)\zeta(t^\prime) \rangle=\sigma^2 \delta(t-t^\prime)$.

\begin{figure}[t!]
  \begin{center}
    \includegraphics[width=5cm,angle=0]{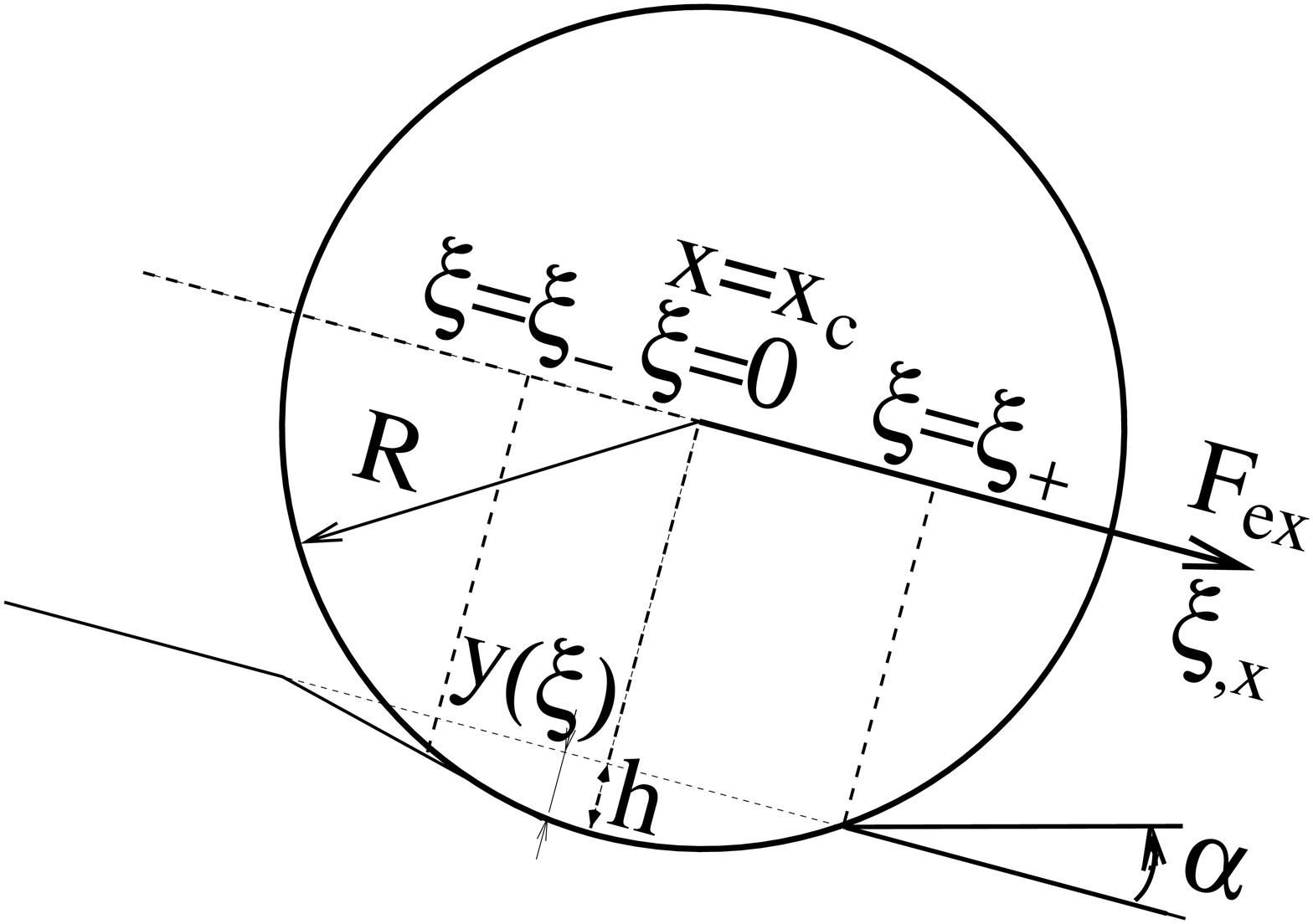}
  \end{center}
  \caption{A rigid cylinder of radius $R$ rolls down an inclined plane. 
    The cylinder contacts the surface of the deformed plane in the
    region $(\xi_-,\xi_+)$ which moves along with the cylinder.  }
  \label{fig1}
\end{figure}

For the rolling friction force $F_R$ we adopt the model developed in
\cite{PSB}, where it is assumed that the plane is composed of springs
of elasticity $k$, mass $m$ and damping constant $\gamma$ to model
elastic, inertial and viscous properties of the surface material. The
rolling friction originates due to the dissipative and inertial
resistance of the springs, which are compressed and subsequently
decompressed when the cylinder rolls on the surface.  As an important
property this simple model leads to a non-monotonous rolling friction
as a function of the velocity \cite{PSB} as observed in experiments
\cite{NonlinFrict}.  Let us briefly sketch the main results of the
model which will be used below. The rolling friction force is given by
\cite{foot}
\begin{equation}
  F_R=-\frac{1}{R}\int_{\xi_-}^{\xi_+}d\xi \xi f(\xi)  +
  m v^2\frac{2h}{R}\,,
  \label{RF}
\end{equation}
where $\xi$ is the coordinate in the comoving frame (Fig. \ref{fig1}),
$ f(\xi)$ is the corresponding force density exerted from the
compressed material on the cylinder,
\begin{equation}
  \label{force:xi}
  f(\xi)=\frac{k}{2R}\xi^2-\frac{\gamma v}{R}\xi+\frac{m v^2}{R}-hk\,,
\end{equation}
and $h(v)$ is the deformation. The force $f(\xi)$ acts in the contact
contact area $\xi_{-} < \xi < \xi_{+}$, where the front boundary
\begin{equation}
  \label{eq:boundaryFront}
\xi_+=\sqrt{2Rh}
\end{equation}
follows from geometry, while for the rear boundary the condition
$f(\xi_-)=0$ yields
\begin{equation}
  \label{eq:boundaryBack}
  \xi_-=\gamma v/k - \sqrt{2hR+\left(\gamma^2/k^2-2 \, m/k \right) v^2}.
\end{equation}
The deformation may be found from the balance of forces normal to the
plane,
\begin{equation}
  Mg \cos \alpha =-\int_{\xi_-}^{\xi_+} f(\xi)  d\xi +\frac{m v^2 }{R}
  \sqrt{2hR}\,.
  \label{hcalc}
\end{equation}
The first term in Eq. \eqref{RF} equals, up to the factor $1/R$, the
total torque exerted on the body due to its interaction with the
deformed surface ($\xi f(\xi)$ is the torque density), while the
second term in Eq. \eqref{RF} as well as in Eq. \eqref{hcalc} accounts
for the surface-cylinder interaction at the {\em infinitesimal} region
at $\xi_+$ where the acceleration diverges as the cylinder rolls over
(for a detailed discussion of this peculiarity see \cite{PSB}). In
this approach it is also assumed that the rolling body does not induce
surface waves.

The viscous drag force due to the surrounding air is given by
\begin{equation}
  \label{air}
  F_D=-Av -Bv^2\,,
\end{equation}
where the (positive) coefficients $A$ and $B$ depend on the Reynolds
number $\mathrm{Re}$, which characterizes the motion of the body.
These coefficients may be estimated from hydrodynamic theory
\cite{LandauLifshHydro}.

For cylinder velocity $\sim 1-10 \, {\rm m/s}$, radius $\sim 0.1-1 \,
{\rm m}$ and kinematic viscosity of the air $\nu = 0.15 \cdot 10^{-4}$
m$^2/s$ the Reynolds number is ${\rm Re} \sim 10^4-10^5$, which
corresponds to the drag force $F_D=\frac12 c \rho v^2 S$ with the
constant $c \approx 0.5$ (here $\rho$ is the air density, $S$ - is the
body cross-section) \cite{LandauLifshHydro}. This yields the estimate
for the constant $B \sim (0.01-1)$ kg/m$^2$.

For very small velocities, i.e. for $\mathrm{Re} \ll 1$, the drag
force is approximately proportional to the velocity.  Lamb's relation
for a cylinder ($F_D=4 \pi \nu \rho v / b$ with $b=\log \left[3.70 \,
  \nu / \left( \mathrm{Re} \,\, v \right) \right]$, $ \mathrm{Re} \, v
/\nu \ll 1$ \cite{LandauLifshHydro}\,) yields the estimate for the
linear constant, $A < 2 \cdot 10^{-4}$ kg/s/m. Thus, for the
parameters addressed here we can neglect the linear drag force, i.e.
$A =0$.  We studied the above model analytically and numerically and
observed the overmentioned effects of bistability and noise controlled
velocity for a wide range of parameters. Here we report the results
for a cylinder of mass per unit length $M=100$ kg/m and radius $R=0.5$
m. The surface material parameters correspond to soft rubber
\cite{matpar}: $m=100$ kg/m$^2$, $k=10^6-10^7$ kg/m$^2$/s$^2$,
$\gamma=10^4-10^5$ kg/m$^2$/s (all parameters per unit length).

{\em Analytical approximation.} As shown below for the given set of
parameters the effects of interest, namely bistability of the rolling
velocity and noise controlled velocity, occur at rather large values
of the velocity. Therefore, as a simplification we assume that the
material does not recover immediately after the cylinder has passed
by, but with a delay, when the cylinder has already left.  The
according approximation $\xi_-=0$ means that the rear part of the
material does not recover on the time scale of the cylinder motion
$\sim R/v$. A similar approximation was investigated in
\cite{GreenwoodMinshallTabor:1961}. The balance equation \eqref{hcalc}
reduces then to $Mg \cos \alpha =\gamma v h +\frac23 \sqrt{2R} k
h^{3/2}$, which yields with the condition of large damping, $v\gamma
\gg k \sqrt{2Rh}$, the velocity dependent deformation
\begin{equation}
  \label{eq:h_a}
  h(v) \approx
\frac{Mg\cos \alpha}{\gamma v+\frac23 k \sqrt{\frac{2RMg \cos \alpha}{ \gamma v}}} \, .
\end{equation}
With the same arguments from Eq. \eqref{RF} we obtain the rolling
friction force
\begin{equation}
  \label{eq:Fr_an}
F_R=\frac{kh^2}{2}+\frac{mv^2h}{R}+\frac{2 \gamma v h}{3 R} \sqrt{2Rh} \, .
\end{equation}

\begin{figure}[t!]
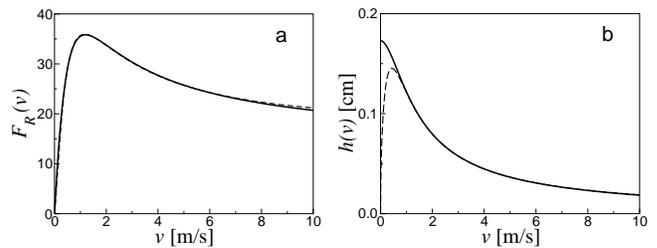

\begin{center}
\includegraphics[width=0.23\textwidth,clip]{FIG2a.eps}
\hspace{0.0cm}
\includegraphics[width=0.23\textwidth,clip]{FIG2b.eps}
\end{center}
\caption{
  a) Rolling friction force $F_R$ and b) deformation $h(v)$ over the
  velocity $v$. The full lines show the numerical solution of Eqs.
  (\ref{eq:boundaryFront},\ref{eq:boundaryBack},\ref{hcalc}) while the
  dashed lines show the approximative theory, Eqs. \eqref{eq:h_a} and
  \eqref{eq:Fr_an}, respectively. Parameters: $m=$100\,kg/m$^2$,
  $k=10^7$\,kg/m$^2$/s$^2$ , $\gamma= 5\times 10^5$\,kg/m$^2$/s,
  $B=0.2$\,kg/m$^2$.  }
\label{fig2}
\end{figure}

{\em Bistability.}  Solving Eqs.
(\ref{eq:boundaryFront},\ref{eq:boundaryBack},\ref{hcalc}) we obtain
the deformation $h(v)$ as a decaying function of the velocity $v$
(Fig. \ref{fig2}b), i.e., the faster the cylinder moves, the less the
surface is compressed.  Using the obtained deformation $h(v)$, the
corresponding rolling friction force may be calculated (Fig.
\ref{fig2}a). $F_R$ depends non-monotonously on the velocity, first it
increases while at larger velocity it decreases.
\begin{figure}[t!]
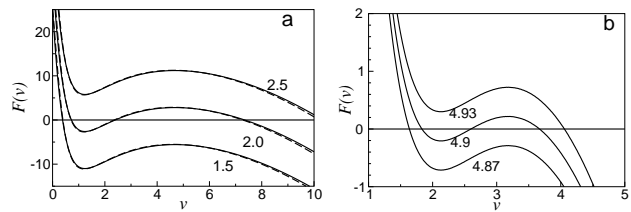

\begin{center}
\includegraphics[height=0.15\textwidth,clip]{FIG3a.eps}
\hspace{0.0cm}
\includegraphics[height=0.15\textwidth,clip]{FIG3b.eps}
\end{center}
\caption{
  Total force $F(v)$ for different angle of inclination (the number
  attached to the curves is the inclination in degrees). Parameters:
  (a) as given in Fig. \ref{fig2}, (b) same, except for $\gamma=
  0.8\times 10^5$\,kg/m$^2$/s, $k=10^6$\,kg/m$^2$/s$^2$,
  $B=0.65$\,kg/m$^2$. Full lines: numerical solution, dashed lines:
  approximative theory.}
\label{figS}
\end{figure}
The non-monotonous behavior of $F_R(v)$ results from the interplay
between the amount of deformed material and the deformation rate: For
small velocity, $F_R(v)$ increases since the surface material is
compressed at increasing rate with growing $v$ while at higher rolling
velocity the amount of deformed material decreases as $h(v)$ decays.
Consequently for fast motion the rolling friction force decays with
increasing velocity.

\begin{figure}[t!]
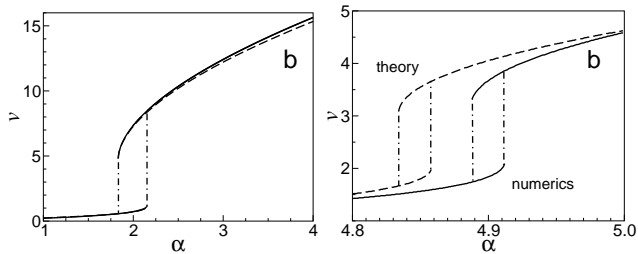

\begin{center}
\includegraphics[width=0.23\textwidth,clip]{FIG4a.eps}
\includegraphics[width=0.23\textwidth,clip]{FIG4b.eps}
\end{center}
\caption{
  Bifurcation diagram for stable velocity states (full line: numerical
  solution, dashed line: analytical approximation).  (a) corresponds
  to Fig. \ref{fig2} and Fig. \ref{figS}a, (b) corresponds to Fig.
  \ref{figS}b.  In both cases the agreement between numerics and
  theory is within few percent.  }
\label{fig3}
\end{figure}

Given the non-monotonous rolling friction force, the total force
acting on the cylinder, ${\cal F}(v)=F_D+F_R+F_{\rm ex}$, is
represented for realistic parameters by a $S$-shaped curve, implying
bistability of the velocity for a certain range of parameters. Indeed,
the steady state condition, $\dot{v}=0$, or ${\cal F}(v)=0$ may be
fulfilled either for only one velocity (top and bottom curves in Figs.
\ref{figS} or for three different velocities (middle curves). The
former case implies a unique stationary velocity, while the later case
allows for three velocities. Only two of them, the smallest and the
largest, correspond to stable motion ($ \partial {\cal F} /\partial v
<0$).  By varying the parameter of the drag force $B$, or by modifying
the driving force $Mg \sin \alpha$ the system may transit from one
stable solution to the other. Figure \ref{fig3} shows the bifurcation
diagram.

\begin{figure}[t!]
\centerline{\includegraphics[width=0.4\textwidth,clip]{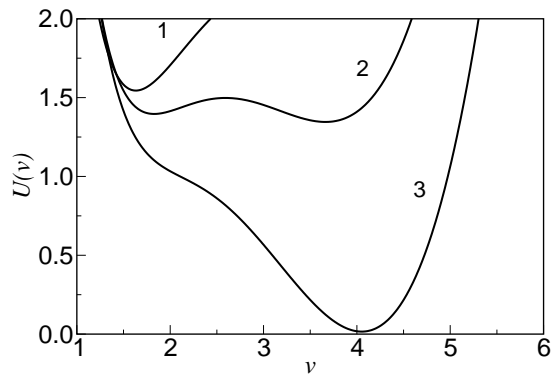}}
\caption{
  Force potential $U(v)$ ($F(v)=-dU/dv$) over the velocity $v$
  (parameters as given in Fig. \ref{fig2}) for angle of inclination
  1$^o$, 2$^o$ and 3$^o$.  }
\label{figPot}
\end{figure}

Starting from an inclination angle that corresponds to a monostable
velocity, the variation of $\alpha$ causes the appearance of the
bistable regime, change of the symmetry of the potential $U(v)$ (Fig.
\ref{figPot}) and the transition to the other monostable regime.

As shown in Figs. \ref{fig2}-\ref{fig3} the analytical theory (dashed
curves) agrees well with the numerical results if the damping
parameter $\gamma$ is large. The deviation in Fig. \ref{fig3} appears
large due to the very detailed resolution of the horizontal axes, in
relative units the deviation is few percent.

\begin{figure}[t!]
%\hspace{-0.2cm}
\includegraphics[width=0.48\textwidth,clip]{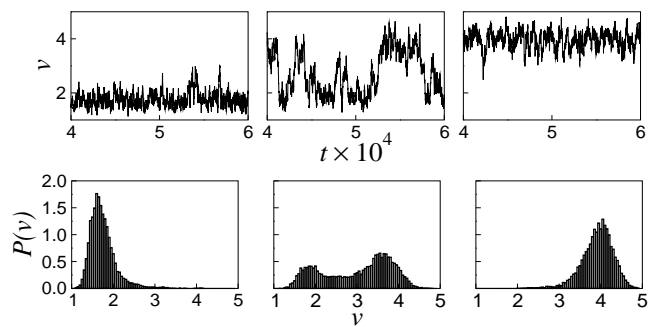}
\caption{Top: velocity of the cylinder subjected to Gaussian white
noise of intensity $\sigma^2$=50 \, kg$^2$ /s$^3$ as obtained from
the numerical integration of Eq. \eqref{Newton}. Bottom: the
corresponding velocity distributions of the cylinder. From left to
right: $\alpha=4.87^o$, $4.9^o$, $4.93^o$, all other parameters
are the same as in Fig. 3b. } \label{model2}
\end{figure}

{\em Noise-induced jumps.}  Up to now we did not consider noise, which
is always present in realistic systems. If the system has only one
stable velocity, the addition of noise does not change the motion of
the cylinder qualitatively. In this case the velocity fluctuates
around the average value given by the steady state condition ${\cal F}
(v)=0$. Figure \ref{model2} (top left and right) shows the velocity of
the cylinder as obtained from a numerical integration of the
stochastic equation \eqref{Newton}. The corresponding velocity
distribution reveals a single peak (Fig. \ref{model2}, bottom left and
right).

For parameters which correspond to bistable velocity the presence of
the noise changes the system behavior qualitatively. Instead of moving
with a fixed velocity there are stochastic jumps between the
meta-stable velocities. Figure \ref{model2} (top, middle) illustrates
this regime. Consequently, the velocity distribution reveals two well
separated peaks (Fig. \ref{model2}, bottom, middle).

\begin{figure}[htbp]
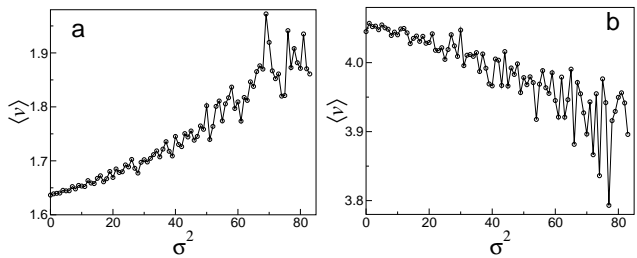

\begin{center}
\includegraphics[width=0.23\textwidth,clip]{FIG7a.eps}
\includegraphics[width=0.23\textwidth,clip]{FIG7b.eps}
\end{center}
\caption{
  Mean velocity of the cylinder over the noise intensity. (a)
  $\alpha=4.87^o$, (b) $\alpha=4.93^o$. The parameters of the system
  are given in Fig. \ref{figS}b.  }
\label{figAccel}
\end{figure}

{\em  Noise-enhanced velocity.} Under certain conditions due to
the nonlinear dependence of the rolling friction on the velocity
the average cylinder velocity may increase with increasing noise
level. Since $v$ increases at  the same driving force $F_{\rm
ex}$, one effectively has increased mobility $\mu=v/F$. This
surprising effect is shown in Fig. \ref{figAccel}.

This phenomenon may be understood analyzing the dynamics of the system
with respect to the force potential $U(v)$. In the absence of noise
the velocity remains in its (meta-)stable state, where $d U/dv =
-{\cal F}(v) = 0$, i.e., at the minimum of the potential $U(v)$ (Fig.
\ref{figPot}). Subjected to the noise, the velocity of the cylinder
fluctuates around this minimum. If the potential $U(v)$ has two
minima, the velocity jumps between both potential wells. If one
potential well of $U(v)$ is narrow while the other is wide, the
average velocity will be shifted towards the wider well. This shift
increases with increasing transition rate between the wells. If the
potential well is wider for the larger stable velocity than for the
smaller one, the average velocity, thus, increases with increasing
noise level (Fig. \ref{figAccel}a). In the opposite case, large noise
impedes rolling (Fig. \ref{figAccel}b).

{\it Summary.}  We investigated the rolling motion of a hard cylinder
on an viscoelastic plane in the presence of a surrounding medium. The
motion of the cylinder is driven by an external force due to the
inclination of the plane. The rolling friction force and the viscous
drag force due to the surrounding medium counteract its motion.  For
certain realistic parameters the stationary velocity of the cylinder
is bistable. The numerical results agree well with the predictions of
a simplified analytical theory in the limit of large damping.

In the presence of noise as it is unavoidable in any realistic
experiment, noise-induced transitions between the meta-stable
velocities are observed. Depending on the system parameters,
increasing noise level may accelerate or decelerate the rolling
motion.

The described effects may be important for technical systems where the
addition of noise may lead to an effective increase of the mobility of
a rolling body driven by an external force.

\end{document}